# Achievable Rate Regions for Dirty Tape Channels and "Joint Writing on Dirty Paper and Dirty Tape"

Reza Khosravi-Farsani, Bahareh Akhbari, Mohammad Reza Aref
Information Systems and Security Lab (ISSL)
Department of Electrical Engineering, Sharif University of Technology, Tehran, Iran
Email: reza.khosravi@ee.sharif.edu, b_akhbari@ee.sharif.edu, aref@sharif.edu

*Abstract*—We consider the Gaussian *Dirty Tape Channel* (DTC) $Y = X + S + Z$, where $S$ is an additive Gaussian interference known causally to the transmitter. The general expression $\max_{P_U, f(\cdot), X=f(U,S)} I(U;Y)$ is presented for the capacity of this channel. For linear assignment to $f(\cdot)$, i.e. $X = U - \beta S$, this expression leads to the *compensation strategy* proposed previously by Willems to obtain an achievable rate for the DTC. We show that linear assignment to $f(\cdot)$ is optimal, under the condition that there exists a real number $\beta^*$ such that the pair $(X + \beta^* S, U)$ is independent of interference $S$. Furthermore, by applying a time-sharing technique to the achievable rate derived by linear assignment to $f(\cdot)$, an improved lower bound on the capacity of DTC is obtained. We also consider the Gaussian multiple access channel with additive interference, and study two different scenarios for this system. In the first case, both transmitters know interference causally while in the second, one transmitter has access to the interference noncausally and the other causally. Achievable rate regions for these two scenarios are then established.

## I. Introduction

Channels with Channel State Information (CSI) available at the transmitter were first studied by Shannon [1]. He investigated a state-dependent discrete channel where the CSI is known causally to the transmitter, and determined its capacity. For this channel with transition probability function $W_{Y|X,S}$, where $Y$ is the channel output, $X$ is the channel input and $S$ is the state of the channel distributed according to $P_S$, Shannon obtained the following expression for the capacity:

$$\mathcal{C} = \max_{P_T, X=T(S)} I(T;Y) \quad (1)$$

where $T$ belongs to the set of all mappings from state alphabet $\mathcal{S}$ to the channel input alphabet $\mathcal{X}$ and is called *strategy*. Therefore, the computation of the capacity requires extension of the channel input alphabet to the space of strategies, and this computation is usually difficult [2].

In another scenario, the transmitter may have access to the CSI noncausally. The capacity of the state-dependent discrete channel where the transmitter knows the CSI noncausally, was obtained by Gelfand and Pinsker in [3]. They showed that the capacity of this channel can be expressed as:

$$\mathcal{C} = \max_{P_{U|S}, X=f(U,S)} I(U;Y) - I(U;S) \quad (2)$$

where $U$ is an auxiliary random variable and $f(\cdot)$ is an arbitrary deterministic function. Costa [4], considered the Gaussian version of the single user channel with noncausal CSI at the transmitter. The channel formulation is given by:

$$Y = X + S + Z \quad (3)$$

where $X$ is the power-constrained input signal, $S$ is the Gaussian interference known noncausally to the transmitter and $Z$ is the unknown Gaussian noise. Costa showed that the capacity of this channel is the same as the case where there is no interference, and

This work was partially supported by Iran National Science Foundation (INSF) under contract No. 84, 5193-2006, and by Iran Telecommunication Research Center (ITRC) under contract No. T500/20958.

is given by $\frac{1}{2}\log\left(1 + \frac{P}{P_z}\right)$, where $P$ is the power associated to the transmitter and $P_z$ is the power of the unknown noise $Z$. In fact, Costa directly treated the Gelfand-Pinsker's channel capacity formula in (2) and showed that by linear assignment to $f(\cdot)$: $X = U - \alpha S$, where $\alpha = \frac{P}{P+P_z}$, for the channel in (3) the rate $\frac{1}{2}\log\left(1 + \frac{P}{P_z}\right)$ is achieved. Due to this remarkable result and Costa's paper title "*writing on dirty paper*," the Gaussian channel in (3), where $S$ is known noncausally to the transmitter, is called *Dirty Paper Channel* (DPC). As a counterpart, the Gaussian channel in (3), where $S$ is known causally to the transmitter is called "*Dirty Tape Channel* (DTC)" [5]. Unlike the DPC, a closed-form formula for the capacity of the Gaussian DTC is still unknown [6]. Willems in [7, 8] considered DTC and presented some techniques to obtain achievable rates for this channel. Specially, using a scheme called "*compensation*" [7], the transmitter expends a part of its power to clean the channel from the known interference noise. Due to the name "*dirty tape channel*", this technique can be referred to as "*partially cleaning dirty tape for writing on it.*" Erez, *et al*, in [5] treated Shannon's capacity formula in (1), and derived some general definitive bounds for the capacity of DTC, and also obtained the *worst-case-interference capacity* in the high Signal-to-Noise Ratio (SNR) regime [5, Theorem 1]. They also obtained a lower bound on the capacity of DTC for general SNR using *inflated lattice strategies* [5, Theorem 2]. It should be noted that the capacity of DTC in the binary case is completely known [9].

In this paper, we first present the following general formula for the capacity of DTC:

$$\mathcal{C}^{DTC} = \max_{\substack{P_U, f(\cdot), X=f(U,S) \\ \mathbb{E}[X^2] \leq P}} I(U;Y) \quad (4)$$

Using this expression, which better shows the relation to the Gelfand-Pinsker's formula (2), would be more convenient for capacity determination of DTC than using Shannon's capacity formula (1). The most significant reason which causes this argument to be subtle, is that capacity determination of DTC via the expression in (4), is essentially equal to obtaining the deterministic function $f(\cdot)$ and the distribution of $U$ which optimize (4). But in Shannon's formula (1) the strategy $T(\cdot)$ itself is a Random Variable (RV) whose alphabet consists of all mappings from $\mathcal{S}$ to $\mathcal{X}$, and since for DTC, $\mathcal{S}$ and $\mathcal{X}$ are the real line, this intuitively makes capacity determination hard.

Similar to what Costa did for the DPC [4] by treating the Gelfand-Pinsker's formula (2), we then examine the expression in (4) for DTC by linear assignment to $f(\cdot)$, i.e., $X = U - \beta S$ and derive an achievable rate for DTC, which leads to the one obtained by compensation strategy proposed by Willems [7]. Using a novel technique, we show that under the condition that there exists a real number $\beta^*$ such that the pair $(X + \beta^* S, U)$ is independent of the state process $S$, the linear assignment to $f(\cdot)$ is optimal. Further, by applying a time-sharing technique to the achievable rate derived by linear assignment to $f(\cdot)$, we obtain an improved lower bound on the capacity of DTC. We also combine (by time-sharing) the latter lower bound and the one derived using inflated lattice strategy in [5], and obtain a new achievable rate for DTC which is very close to the trivial upper bound $\frac{1}{2}\log\left(1+\frac{P}{P_z}\right)$ in all SNR values.

Moreover, it is interesting to note that although Costa's result for DPC was generalized to some multiuser settings in [10, 11, 12], up to now there is no result for the multiuser DTCs (not even an example about it). In this paper, we show that our approach of obtaining the achievable rates for the DTC using the expression in (4) and linear assignment to $f(\cdot)$, can be generalized to multiuser systems and hence achievable rate regions can be obtained for the multiuser DTCs. To this end, we consider a two-user Gaussian Multiple Access Channel (MAC) with additive Gaussian interference. Two different scenarios are studied: in the first scenario both transmitters know the interference $S$ causally, while in the second scenario, one transmitter has access to the interference noncausally and the other has access causally, where we call it "*joint writing on dirty paper and dirty tape*." Then, we establish achievable rate regions for these scenarios.

The rest of the paper is organized as follows: In Section II channel model definitions are given, and in Section III, the main results are presented.

## II. CHANNEL MODELS AND DEFINITIONS

Because of space limitations, we deal with channel model definitions only for MAC with CSI. Clearly, definitions for the single user channel with CSI can be obtained simply from those of MAC with CSI.

*Definition 1:* A two-user discrete memoryless MAC with CSI, denoted by $\{\mathcal{X}_1, \mathcal{X}_2, \mathcal{Y}, \mathcal{S}, P_S(s), W(y|x_1, x_2, s)\}$ is a channel with input alphabets $\mathcal{X}_1, \mathcal{X}_2$ and output alphabet $\mathcal{Y}$. The transition probability function $W(y|x_1, x_2, s)$ describes the relation between channel inputs, channel state, and channel output. We also assume that the state process which takes values over the alphabet $\mathcal{S}$ is independently identically distributed (i.i.d), and drawn according to a known probability mass function (p.m.f) $P_S(s)$. For discrete channel, all alphabets are finite sets. The Gaussian (continuous) version of this channel is given by:

$$Y = X_1 + X_2 + S + Z \quad (5)$$

where $X_i$ is the real-valued input signal at transmitter-$i$, $i = 1,2$, $Y$ is the real-valued output signal at the receiver, and $Z \sim \mathcal{N}(0, P_z)$ is the zero mean (unknown) i.i.d Gaussian noise. The additive interference noise $S$, which is known causally or noncausally to the transmitters, is also assumed to be i.i.d Gaussian with power $P_s$: $S \sim \mathcal{N}(0, P_s)$, and independent of the unknown noise $Z$. Transmitter-$i$ is subject to an average power constraint $P_i$, as: $\mathbb{E}[X_i^2] \leq P_i, i = 1,2$.

*Definition 2:* A length-$n$ block code $\mathfrak{C}(n, R_1, R_2)$ with rate $(R_1, R_2)$ for the state-dependent MAC consists of: a) two message sets $\mathcal{W}_i = \{1, ..., 2^{nR_i}\}, i = 1,2$ where two messages $W_1, W_2$ are distributed (independent of each other) uniformly over respective sets, b) two sets of encoding functions $\{\xi_{i,t}(\cdot)\}_{t=1}^n, i = 1,2$, and c) a decoding function $\Lambda(\cdot)$:

*Encoding and decoding:* In terms that each transmitter has access to the state process causally or noncausally, three different situations may arise:

1) Both transmitters have access to CSI noncausally. In this case, $\xi_{i,t}(\cdot), i = 1,2, t = 1, ..., n$, is given by: $\xi_{i,t}: \mathcal{W}_i \times \mathcal{S}^n \to \mathcal{X}_i$ which generates $X_{i,t} = \xi_{i,t}(W_i, S^n)$.

2) Both transmitters have access to CSI causally. In this case, $\xi_{i,t}(\cdot), i = 1,2, t = 1, ..., n$, is given by: $\xi_{i,t}: \mathcal{W}_i \times \mathcal{S}^t \to \mathcal{X}_i$ which generates $X_{i,t} = \xi_{i,t}(W_i, S^t)$.

3) One transmitter knows CSI noncausally and the other knows causally. In this case at each time instant $t, t = 1, .., n$, the encoding function for transmitter-1, i.e., $\xi_{1,t}(\cdot)$, is given by: $\xi_{1,t}: \mathcal{W}_1 \times \mathcal{S}^n \to \mathcal{X}_1$ which generates $X_{1,t} = \xi_{1,t}(W_1, S^n)$, while for transmitter-2, i.e., $\xi_{2,t}(\cdot)$ is given by: $\xi_{2,t}: \mathcal{W}_2 \times \mathcal{S}^t \to \mathcal{X}_2$ which generates $X_{2,t} = \xi_{2,t}(W_2, S^t)$.

We also assume that the receiver has no knowledge of CSI, so disregarding the causality of CSI at the transmitters, the decoder $\Lambda(\cdot)$ is given by: $\Lambda: \mathcal{Y}^n \to \mathcal{W}_1 \times \mathcal{W}_2$ which estimates the messages as $(\widehat{W}_1, \widehat{W}_2) = \Lambda(Y^n)$.

The error probability of the code and also the capacity for the presented channel models are defined as usual, so details are omitted here. Finally, it should be noted that the Gaussian MAC in (5), where both transmitters have access to CSI noncausally, stands for "*writing on dirty paper*" (this situation was investigated in [10], and we do not consider it here). Moreover, the case where both transmitters have access to CSI causally stands for "*writing on dirty tape*," and the case where one transmitter has access to CSI noncausally and the other causally, stands for "*joint writing on dirty paper and dirty tape*."

## III. MAIN RESULTS

We state our results for the single user and multiple access DTCs in Section III-A and III-B, respectively. The scenario of joint writing on DPC and DTC is addressed in Section III-C.

*III-A) The single user writing on DTC:*

At first, we present the general formula given in (4) for the capacity of DTC, in the following:

*Observation:* The capacity of the DTC (3) in which the channel state $S$ is known causally to the transmitter, is given by (4).

*Proof:* The proof of achievability is the same as Shannon's strategy [1], however here the channel input is given by $X = f(U, S)$, while in the Shannon's approach that is of the form $X = T(S)$, where $T$ is the *strategy*. For the converse part, using Fano's inequality, we can write:

$$nR - \epsilon_n \leq I(W; Y^n) = \sum_{t=1}^{n} I(W; Y_t | Y^{t-1})$$
$$\leq \sum_{t=1}^{n} I(W, S^{t-1}, Y^{t-1}; Y_t) = \sum_{t=1}^{n} I(\widehat{U}_t; Y_t) \quad (6)$$

where $\epsilon_n \to 0$ as $n \to \infty$, and $\widehat{U}_t \triangleq (W, S^{t-1}, Y^{t-1})$. Now, define a time-sharing RV $Q$, uniformly distributed on $\{1, ..., n\}$ and independent of everything. Set $U \triangleq (\widehat{U}_Q, Q), X \triangleq X_Q, S \triangleq S_Q$ and $Y \triangleq Y_Q$, then from (6), we have: $R \leq I(U; Y)$. On the other hand, by definition of causal encoding, $H(X_t | \widehat{U}_t, S_t) = 0$. So, there exists a deterministic function $f(\cdot)$ such that $X_Q = f(\widehat{U}_Q, Q, S_Q)$. Furthermore, one can easily verify that the power constraint at the transmitter implies $\mathbb{E}[X^2] = \mathbb{E}[X_Q^2] \leq P$. ∎

*Remark:* For the capacity of *discrete memoryless* channel with causal CSI at the transmitter, a same formula as in (4) was previously reported in [13]. Note that for the discrete channel the cardinality of $U$ is bounded above as $\|\mathcal{U}\| \leq \|\mathcal{Y}\|$, however for the Gaussian DTC there is no restriction on the auxiliary RV $U$.

As mentioned before, by the general formula in (4), capacity determination for DTC reduces essentially to obtaining the deterministic function $f(\cdot)$ and distribution of $U$ which optimize (4). Now, we examine the expression in (4) by linear assignment to $f(\cdot)$, together with Gaussian assignment to $U$, and derive an achievable rate for DTC.

Assume that $\beta \in [0, \sqrt{\frac{P}{P_s}}]$, be an arbitrary real number. Let $U$ be a Gaussian distributed RV such that $U \sim \mathcal{N}(0, P - \beta^2 P_s)$ and be independent of $S$. Now, by setting $X \triangleq U - \beta S$ in (4), we obtain:

$$I(U; Y) = h(X + S + Z) - h(X + S + Z | U)$$
$$= \frac{1}{2}\log\left(1 + \frac{P - \beta^2 P_s}{P_z + (1-\beta)^2 P_s}\right) \quad (7)$$

where $h(\cdot)$ denotes the differential entropy. Now, denote:

$$C_1(P) = \max_{\beta \in [0, \sqrt{\frac{P}{P_s}}]} \frac{1}{2}\log\left(1 + \frac{P - \beta^2 P_s}{P_z + (1-\beta)^2 P_s}\right) \quad (8)$$

So, we obtain that $C_1(P)$ is achievable for DTC, i.e., $C^{DTC} \geq C_1(P)$. Further, one can see that (8) is optimized when $\beta = \frac{P + P_z + P_s - \sqrt{(P + P_z + P_s)^2 - 4PP_s}}{2P_s}$. In fact to achieve the rate $C_1(P)$, the transmitter expends a part $\beta^2 P_s$ of its power to clean the dirty tape and uses the rest of its power to transmit the message.

As we see, linear assignment to $f(\cdot)$ in (4), together with Gaussian assignment to $U$, leads to the achievable rate previously derived for DTC by using the compensation strategy in [7]. Now, two questions arise naturally: First, can linear assignment to $f(\cdot)$ together with any other distribution on the RV $U$, lead to achievable rates higher than $C_1(P)$ in (8) for DTC? In the following, we answer to this:

*Theorem 1:* The expression in (4), for linear assignment to $f(\cdot)$, i.e., $X = U - \beta S$ where $\beta$ is a real number and $U$ is a real-valued RV, is always less than $C_1(P)$ given by (8), disregarding any distribution on the RV $U$.

*Proof:* By substituting $X = U - \beta S$, in (4), we have:
$$I(U;Y) = h(X + S + Z) - h(X + S + Z|U)$$
$$= h(U - \beta S + S + Z) - h(U - \beta S + S + Z|U)$$
$$\stackrel{(a)}{=} h(U + (1-\beta)S + Z) - h((1-\beta)S + Z)$$
$$\stackrel{(b)}{\leq} \frac{1}{2}\log\left(1 + \frac{P - \beta^2 P_s}{P_z + (1-\beta)^2 P_s}\right) \leq C_1(P) \quad (9)$$

where the equality (a) is due to the independence of $U$ from $S$ and $Z$, and the inequality (b) is due to the Maximum Entropy Theorem [14, Theorem 8.6.5]. ∎

Another question is whether the linear assignment to $f(\cdot)$ is optimal for the capacity expression given in (4)? The answer is no, and in fact later we will show that some improved lower bounds on the capacity of DTC can be obtained. Nevertheless, in the next theorem, using a novel idea, we show that when we restrict our attention to the case where the triple $(X, S, U)$ is assumed to be jointly Gaussian, through using general formula given in (4) no achievable rate higher than $C_1(P)$ in (8) can be obtained for DTC.

*Theorem 2:* Under condition that there exists a real number $\beta^*$, such that the pair $(X + \beta^* S, U)$ is independent of the state process $S$, the expression in (4) for any arbitrary assignment to $f(\cdot)$ (linear or nonlinear), is always less than $C_1(P)$ in (8).

*Proof:* For the expression in (4), we can write:
$$I(U;Y) = h(X + S + Z) - h(X + S + Z|U) \quad (10)$$

Since $X + \beta^* S$ is independent of $S$, the parameter $\beta^*$ should satisfy:
$$\beta^* = -\frac{\mathbb{E}[XS]}{P_s} \quad (11)$$

So, due to Maximum Entropy Theorem:
$$h(X + S + Z) \leq \frac{1}{2}\log(2\pi e(P + P_s + P_z - 2\beta^* P_s)) \quad (12)$$

For the second term in (10) we have:
$$h(X + S + Z|U) = h(X + \beta^* S + (1-\beta^*)S + Z|U) \quad (13)$$

Note that since according to the hypothesis, $Z$ and $S$ are independent of the pair $(X + \beta^* S, U)$, the linear combination $(1-\beta^*)S + Z$ is independent of $(X + \beta^* S, U)$, too. Now, using the *entropy power inequality* [15, Lemma 2], we have:
$$h(X + \beta^* S + (1-\beta^*)S + Z|U)$$
$$\geq \log\left(e^{h(X+\beta^* S|U)} + e^{h((1-\beta^*)S+Z|U)}\right)$$
$$\geq \log\left(e^{h((1-\beta^*)S+Z|U)}\right)$$
$$= h((1-\beta^*)S + Z|U) = h((1-\beta^*)S + Z)$$
$$= \frac{1}{2}\log(2\pi e((1-\beta^*)^2 P_s + P_z)) \quad (14)$$

Substituting (12)-(14) into (10), we obtain:
$$I(U;Y) \leq \frac{1}{2}\log\left(1 + \frac{P - \beta^{*2} P_s}{P_z + (1-\beta^*)^2 P_s}\right) \leq C_1(P) \quad (15)$$

This completes the proof. ∎

*Remarks:*

1) Note that if we impose that the triple $(X, S, U)$ should be jointly Gaussian distributed, then it is easy to verify that the condition in Theorem 2 is satisfied (in fact it is enough to set $\beta^*$ as in (11)). So, we conclude that the expression in (4) for the jointly Gaussian variables, disregarding any assignment to $f(\cdot)$ (linear or not), does not lead to any achievable rates higher than $C_1(P)$ for DTC.

2) Theorem 1, together with Theorem 2 show that to obtain achievable rates higher than $C_1(P)$ for DTC by examining (4), one should consider nonlinear assignments to $f(\cdot)$, such that the triple $(X, S, U)$ would not be jointly Gaussian. However, computation of (4) under these conditions is usually difficult.

As been pointed out in [7], the achievable rate $C_1(P)$ given in (8) is not a concave function of $P$ in general, and one can improve it using time-sharing technique. We deal with this, in the next proposition:

*Proposition 1:* Denote:
$$C_2(P) = \max_{0 \leq \lambda, \xi \leq 1}\left(\lambda C_1\left(\xi \frac{P}{\lambda}\right) + \bar{\lambda} C_1\left(\bar{\xi} \frac{P}{\bar{\lambda}}\right)\right) \quad (16)$$

where $\bar{x} \triangleq 1 - x$, for $0 \leq x \leq 1$. Then the capacity of DTC is lower bounded by $C^{DTC} \geq C_2(P)$.

*Proof:* To achieve $C_2(P)$ in (16), two modes of operation are considered: for a fraction $\lambda$ of the time, the transmitter uses a fraction $\xi$ of its power (the transmitted signals are with power $\xi \frac{P}{\lambda}$) for transmission, and during the remaining fraction $\bar{\lambda}$ of the time, it applies the rest of the power, i.e., $\bar{\xi} P$, for transmission (transmission for both modes includes cleaning the dirty tape and also transmitting the message.) ∎

*Remark:* One can easily check that $C_2(P)$ is a concave function of $P$.

Now, we mention the achievable rate derived in [5] for DTC using inflated lattice strategy. Denote:
$$C_3(P) = \max\left\{0, \frac{1}{2}\log\left(1 + \frac{P}{P_z}\right) - \frac{1}{2}\log\left(\frac{2\pi e}{12}\right)\right\} \quad (17)$$

*Lemma 1* [5, Theorem 2]: The capacity of DTC is lower bounded as $C^{DTC} \geq C_3(P)$.

In Fig. 1 the achievable rate $C_2(P)$ and also $C_3(P)$ have been plotted. The trivial upper bound on the capacity of DTC, i.e., $\frac{1}{2}\log\left(1 + \frac{P}{P_z}\right)$ has also been plotted. It is clear from the figure that for low and intermediate values of SNR, $C_3(P)$ is higher than $C_2(P)$, while for high SNRs, $C_2(P)$ is higher than $C_3(P)$. It should be noted that the achievable rate $C_2(P)$ is arbitrarily close to the trivial upper bound, provided the value of SNR, i.e. $\frac{P}{P_z}$ is sufficiently large. Also note that the achievable rate $C_3(P)$ is not a concave function of $P$. So, by applying a time-sharing technique between these two latter rates, a new achievable rate can be obtained for the capacity of DTC. Denote:
$$C_4(P) = \max_{0 \leq \lambda, \xi \leq 1}\left(\lambda C_2\left(\xi \frac{P}{\lambda}\right) + \bar{\lambda} C_3\left(\bar{\xi} \frac{P}{\bar{\lambda}}\right)\right) \quad (18)$$

*Proposition 2:* The rate $C_4(P)$ in (18) is achievable for DTC.

*Proof:* The proof is the same as Proposition 1. However, now in the first mode, for a fraction $\lambda$ of time the transmitter applies the time-sharing-based compensation strategy (the scheme for achieving $C_2(P)$) and expends a fraction $\xi$ of its power for transmission, while in the second mode, during the remaining fraction $\bar{\lambda}$ of the time, the transmitter applies the inflated lattice strategy [5] and uses the rest of its power, i.e., $\bar{\xi} P$, for transmission. ∎

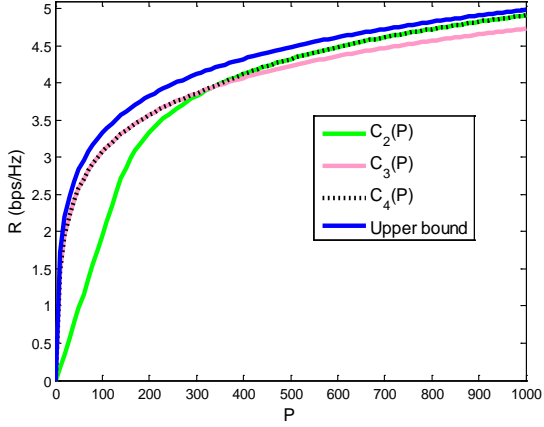

Figure 1. Achievable rates for DTC: $P_s = 100$, $P_z = 1$.

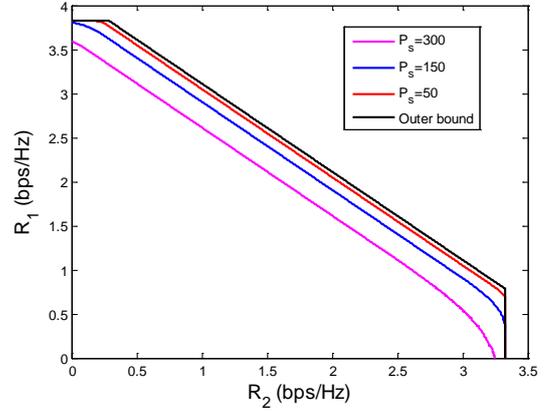

Figure 2. Achievable rate regions for multiple access DTC: $P_1 = 2P_2 = 200$, $P_z = 1$.

We also have plotted the achievable rate $C_4(P)$ in Fig. 1. As it is clear from the figure, the latter achievable rate for DTC ($C_4(P)$) is higher than $C_2(P)$ and $C_3(P)$ for all values of SNRs. Further, it is very close to the trivial upper bound.

*III-B) Multiple access writing on DTC:*

In this section, we extend the approach of "*partially cleaning dirty tape for writing on it*" to the multiple access dirty tape channel and derive an achievable rate region for it. First, we recall an achievable rate region derived in [13] for the two-user discrete memoryless MAC with causal CSI at both transmitters:

*Lemma 2* [13]: For the two-user discrete memoryless MAC with causal CSI at both transmitters, the following rate region is achievable:

$$\bigcup_{\substack{P_{U_1} \times P_{U_2} \\ X_i = f_i(U_i, S), i=1,2}} \begin{Bmatrix} 0 \leq R_1, R_2 \\ R_1 \leq I(U_1; Y|U_2) \\ R_2 \leq I(U_2; Y|U_1) \\ R_1 + R_2 \leq I(U_1, U_2; Y) \end{Bmatrix} \quad (19)$$

Now, consider the multiple access DTC defined by (5), where the state noise $S$ is causally known to both transmitters and transmitter-$i$ is subject to power constraint $P_i$, $i = 1,2$. It can be easily checked that the same rate region as the one given in (19) under power constraints $\mathbb{E}[X_i^2] \leq P_i, i = 1,2$, is achievable for the Gaussian multiple access DTC. Therefore, we derive an achievable rate region for this channel as follows.

Assume that $\beta_i \in [0, \sqrt{\frac{P_i}{P_s}}]$, $i = 1,2$, be two arbitrary real numbers. Define two Gaussian distributed RVs $U_i$, $i = 1,2$, independent of each other and also independent of the state $S$, such that $U_i \sim \mathcal{N}(0, P_i - \beta_i^2 P_s)$. Denote $X_i \triangleq U_i - \beta_i S$, $i = 1,2$. Now, using these parameters to evaluate (19), we have the following theorem:

*Theorem 3:* For the two-user multiple access DTC in (5), with causal CSI at both transmitters, the following rate region is achievable:

$$\bigcup_{\substack{\beta_1, \beta_2 \\ \beta_i \in [0, \sqrt{\frac{P_i}{P_s}}], i=1,2}} \begin{Bmatrix} 0 \leq R_1, R_2 \\ R_1 \leq \frac{1}{2}\log\left(1 + \frac{P_1 - \beta_1^2 P_s}{(1 - \beta_1 - \beta_2)^2 P_s + P_z}\right) \\ R_2 \leq \frac{1}{2}\log\left(1 + \frac{P_2 - \beta_2^2 P_s}{(1 - \beta_1 - \beta_2)^2 P_s + P_z}\right) \\ R_1 + R_2 \leq \frac{1}{2}\log\left(1 + \frac{P_1 + P_2 - (\beta_1^2 + \beta_2^2)P_s}{(1 - \beta_1 - \beta_2)^2 P_s + P_z}\right) \end{Bmatrix}$$
(20)

In fact, to achieve the rate region in (20) for the multiple access DTC, transmitter-$i$, $i = 1,2$, expends a part $\beta_i^2 P_s$ of its power to clean the channel from the known state noise $S$, and uses the rest of its power, i.e., $P_i - \beta_i^2 P_s$ to transmit the message intended to it.

In Fig. 2, the achievable rate region in (20) has been plotted for different values of $P_s$. We have also plotted the trivial upper bound, i.e., the capacity region of the Gaussian MAC (state-independent). As Fig. 2 shows, when $P_s$ is lower than one of the associated powers to transmitters ($P_1$ or $P_2$), some boundary points of the trivial upper bound are also achievable. In fact, at these points, one transmitter cleans the known noise from the channel completely, so the other transmitter can transmit at its capacity rate, i.e., $\frac{1}{2}\log\left(1 + \frac{P_i}{P_z}\right)$, where $i = 1$ or $2$.

Note that the approach followed here can also be treated for other multiuser DTCs. Therefore, achievable rate regions for the *broadcast* and *relay* DTCs can be established in the same way, which have reported in [16].

*III-C) Joint writing on DPC and DTC:*

In this section, we study the scenario of joint writing on DPC and DTC, i.e., a Gaussian two-user MAC with CSI defined in (5), where one transmitter has access to the state noncausally and the other causally. In fact, the channel from the viewpoint of the first user is a dirty paper while from the viewpoint of the second user is a dirty tape. We establish an achievable rate region for this scenario. First, we consider the discrete memoryless counterpart of this scenario in the next theorem:

*Theorem 4:* For the discrete memoryless two-user MAC with CSI, where the first transmitter knows the CSI noncausally and the second knows it causally, the following rate region is achievable:

$$\bigcup_{\substack{P_{U_1|S} \times P_{U_2} \\ X_i = f_i(U_i, S), i=1,2}} \begin{Bmatrix} 0 \leq R_1, R_2 \\ R_1 \leq I(U_1; Y|U_2) - I(U_1; S) \\ R_2 \leq I(U_2; Y|U_1) \\ R_1 + R_2 \leq I(U_1, U_2; Y) - I(U_1; S) \end{Bmatrix} \quad (21)$$

*Proof:* The proof of achievability of the above rate region is nearly the same as [17, Theorem 1], so we only sketch an outline of the proof. Encoding and decoding strategy is as follows. Fix p.m.fs $P_{U_1|S}(u_1|s)$, $P_{U_2}(u_2)$, and also two deterministic functions $f_i(\cdot): \mathcal{U}_i \times \mathcal{S} \to \mathcal{X}_i, i = 1,2$. Compute the distribution $P_{U_1}(u_1)$ as $P_{U_1}(u_1) = \sum_s P_S(s) P_{U_1|S}(u_1|s)$. At the first transmitter, for each $w_1 \in \{1, \ldots, 2^{nR_1}\}$, generate $2^{nR_0}$ codewords $\{u_1^n(w_1, j): j = 1, \ldots, 2^{nR_0}\}$ at random according to $\prod_{t=1}^n p(u_{1,t})$, forming the bin corresponding to $w_1$. Therefore, $u_1^n(w_1, j)$ represents a codeword indexed by $w_1 \in \{1, \ldots, 2^{nR_1}\}$ which is the bin index, and $j \in [1, 2^{nR_0}]$ which is an index within the bin. At the second transmitter, for each $w_2 \in \{1, \ldots, 2^{nR_2}\}$, generate at random a codeword $u_2^n(w_2)$ according to $\prod_{t=1}^n p(u_{2,t})$. This codebook is revealed to both transmitters and also to the receiver. The transmitter-1 which wants to send the message $w_1 \in \{1, \ldots, 2^{nR_1}\}$, knowing the sequence $S^n = s^n$, seeks a codeword in the bin $w_1$ that is jointly strongly typical [14, Sec. 10.6] with $s^n$, denoted as $u_1^n(w_1, j)$. If multiple such codewords in the bin $w_1$ exist, transmitter-1 chooses the one with smallest $j$. If no such index $j$ exists, then an encoding error is declared at the first encoder. The transmitter-1 then sends $x_{1,t} = f_1(u_{1,t}(w_1, j), s_t)$, $t = 1, \ldots, n$, over the channel. To send the message $w_2 \in \{1, \ldots, 2^{nR_2}\}$, transmitter-2 knowing $S_t = s_t$, sends $x_{2,t} = f_2(u_{2,t}(w_2), s_t)$, $t = 1, \ldots, n$, over the channel. The decoder receives $Y^n = y^n$ according to $\prod_{t=1}^n W(y_t|x_{1,t}, x_{2,t}, s_t)$, and declares that $(\hat{w}_1, \hat{w}_2)$ have been sent if there are unique indices such that $(u_1^n(\hat{w}_1, j), u_2^n(\hat{w}_2))$ is jointly strongly typical with $y^n$ for some $j \in \{1, \ldots, 2^{nR_0}\}$. Thus $(\hat{w}_1, \hat{w}_2) = (w_1, w_2)$ with arbitrarily high probability for large enough $n$, if:

$$\begin{cases} R_0 + R_1 \leq I(U_1; Y|U_2) \\ R_2 \leq I(U_2; Y|U_1) \\ R_0 + R_1 + R_2 \leq I(U_1, U_2; Y) \end{cases} \quad (22)$$

Moreover, one can see that the error probability of encoding at transmitter-1 tends to zero if $R_0 > I(U_1; S)$. Therefore, by applying a Fourier-Motzkin elimination to remove $R_0$ from (22), it yields that any rate pair $(R_1, R_2)$ which belongs to the rate region given in (21) is achievable. This completes the proof. ∎

Now, consider a Gaussian version of the underlying channel, where transmitter-$i$ is subject to power constraint $P_i$, $i = 1, 2$. Again, one can see that the same rate region as (21) under power constraints $\mathbb{E}[X_i^2] \leq P_i, i = 1, 2$, is achievable for the Gaussian multiple access joint writing on dirty paper and dirty tape channel. By this observation, we state our achievable rate region for this channel in the following theorem:

*Theorem 5:* For the Gaussian two-user MAC in (5), where the first user knows the state process noncausally and the second user knows it causally, i.e., joint writing on dirty paper and dirty tape channel, the following rate region is achievable:

$$\bigcup_{\substack{\alpha \in \mathbb{R}, \\ \beta \in \left[-\sqrt{\frac{P_2}{P_s}}, \sqrt{\frac{P_2}{P_s}}\right]}} \begin{cases} 0 \leq R_1, R_2 \\ R_1 \leq \frac{1}{2}\log\left(\frac{P_1 + (1-\beta)^2 P_s + P_z}{(1-\alpha-\beta)^2 P_s + \alpha^2 \frac{P_s}{P_1} P_z + P_z}\right) \\ R_2 \leq \frac{1}{2}\log\left(1 + \frac{(P_2 - \beta^2 P_s)\left(1 + \alpha^2 \frac{P_s}{P_1}\right)}{(1-\alpha-\beta)^2 P_s + \alpha^2 \frac{P_s}{P_1} P_z + P_z}\right) \\ R_1 + R_2 \leq \frac{1}{2}\log\left(\frac{P_1 + P_2 + (1-2\beta)P_s + P_z}{(1-\alpha-\beta)^2 P_s + \alpha^2 \frac{P_s}{P_1} P_z + P_z}\right) \end{cases} \quad (23)$$

*Proof:* Assume that $\alpha \in \mathbb{R}$, and $\beta \in \left[-\sqrt{\frac{P_2}{P_s}}, \sqrt{\frac{P_2}{P_s}}\right]$ are two arbitrary real numbers. Let $U_1 \sim \mathcal{N}(0, P_{U_1})$ be a RV which is jointly Gaussian with state $S$, such that:

$$P_{U_1} \triangleq \mathbb{E}[U_1^2] = P_1 + \alpha^2 P_s, \quad \mathbb{E}[U_1 S] = \alpha P_s \quad (24)$$

Also let $U_2 \sim \mathcal{N}(0, P_2 - \beta^2 P_s)$ be a Gaussian RV independent of the pair $(U_1, S)$. Define $X_1$ and $X_2$ as:

$$X_1 \triangleq U_1 - \alpha S \qquad X_2 \triangleq U_2 - \beta S \quad (25)$$

Note that (24) and (25) imply $\mathbb{E}[X_1 S] = 0$, and since $X_1$ and $S$ are jointly Gaussian, they are independent. Now, using $(U_1, U_2, X_1, X_2)$ as defined by (24) and (25) to evaluate (21), we obtain the achievability of (23). ∎

In fact, to achieve the rate region in (23) for joint writing on dirty paper and dirty tape channel, the first transmitter uses Costa's encoding [4], while the second user applies the compensation strategy and expends a fraction of its power to clean the channel from the known noise.

In Fig. 3, we have plotted the rate region given in (23), for different values of $P_s$, ($P_1, P_2$ and $P_z$ are same as in Fig. 2), and also the trivial upper bound, i.e., the capacity region of the Gaussian MAC (state-independent).

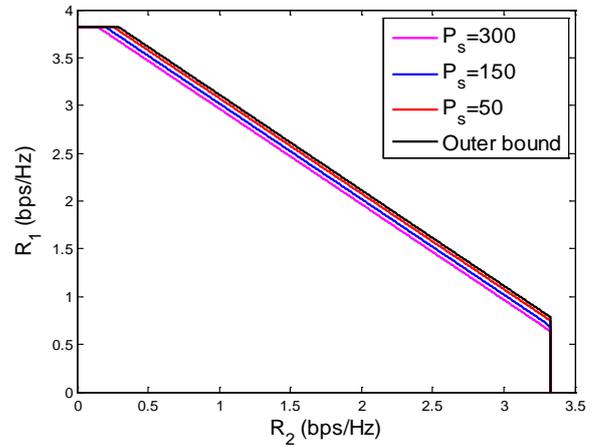

Figure 3.  Achievable rate regions for joint writing on DPC and DTC: $P_1 = 2P_2 = 200, P_z = 1$.

Comparing Fig. 2 and Fig. 3, it is clear that the inner bound on the capacity region of joint writing on DTC and DPC scenario depicted in Fig. 3, is much closer to the trivial upper bound than that is depicted in Fig. 2 which is for the previous setup. Finally, notice that one can improve the achievable rate region in (23), provided that the first encoder uses *Generalized Dirty Paper Coding* (GDPC) [17, Section 4.1], and the second encoder applies the compensation strategy. This improved achievable rate region is reported in [16].